\documentclass[sn-mathphys]{sn-jnl}

\usepackage{longtable}
\usepackage{caption}
\jyear{2021}%

\theoremstyle{thmstyleone}%
\theoremstyle{thmstyletwo}%

\theoremstyle{thmstylethree}%

\begin{document}
\title[Monu et al.]{Analysis of the Impact of North Indian Ocean Cyclonic Disturbance on Human and Economic Losses}


\author[1]{\fnm{Monu} \sur{Yadav}}\email{yadavm012@gmail.com}

\author*[1]{\fnm{Laxminarayan} \sur{Das}}\email{lndas@dce.ac.in}


\affil*[1]{\orgdiv{Department of Applied Mathematics}, \orgname{Delhi Technological University}, \orgaddress{\city{New Delhi}, \country{India}}}




\abstract{This paper explores the features of cyclonic disturbances (CDs) in the North Indian Ocean (NIO) by utilizing data from 1990 to 2022. It investigates the occurrence rate of these disturbances and their effects on human and economic losses throughout the mentioned period. The analysis demonstrates a rising trend in the occurrence of CDs in the NIO. While there has been a slight decline in CD-related fatalities since 2015, but there has been a considerable increase in economic losses. These findings can be attributed to enhanced government initiatives in disaster prevention and mitigation in recent years, as well as rapid economic growth in regions prone to CDs. The study sheds light on the significance of addressing the impact of CDs on both human lives and economic stability in the NIO region.}





\maketitle

\section{Introduction}\label{sec1}

Cyclonic disturbances (CDs) pose a significant hazadies challenge to the global community, resulting in substantial human and economic losses. These disturbances manifest as intense low-pressure areas within the Earth's atmospheric system, leading to extreme weather events. Once these disturbances exceed the minimum threshold of 34 knots (18 meters per second), they are universally known as `` Tropical cyclone" or simply ``cyclone," as defined by the World Meteorological Organization (WMO) \cite{bib1}. The classification of these weather systems mainly revolves around their wind speed and applies to  regions characterized by strong winds. W.M. Gray \cite{bib14} identified six features essential for the development of large-scale CDs, which, although necessary, are not sufficient conditions for their formation. These features are:
\begin{itemize}
	\item Sea surface temperature energy [SST $>$ 26°C to a depth of 60 m].
	\item Enhanced mid-troposphere (700 hPa) relative humidity.
	\item Conditional instability.
	\item Enhanced lower troposphere relative vorticity.
	\item Weak vertical shear of the horizontal winds at the genesis site.
	\item Displacement by at least 5° latitude away from the equator.
\end{itemize}

While CDs are known by various names globally, such as  tropical cyclone, typhoons, and hurricanes, they share common characteristics. The primary difference lies in the names assigned to them in different regions. It is crucial to note that the classification of CDs varies from region to region, reflecting the specific criteria and thresholds set by meteorological organizations in each area. Regional Specialized Meteorological Centers (RSMCs) determine the classification of different phases of CDs based on their intensity, employing customized systems. In the North Indian Ocean (NIO) region, the Indian Meteorological Department (IMD) serves as the RSMC and has developed its own classification system, as shown in Table 1 (adapted from Table 1 of the referenced paper \cite{bib2}).

\begin{table}[ht]
	\centering
	\begin{tabular}{|c|c|c|}
		\hline
		Phases& T Number& Maximum Wind Speed  \\
		\hline
		Low pressure area (WML)& T1.0 & $<$ 17 kts   \\
		
		Depression (D) & T1.5 & 17-27 kts  \\
		
		Deep Depression (DD) & T2.0 & 28-33 kts  \\
		
		Cyclonic Storm (CS) & T2.5-T3.0 & 34-47 kts  \\
		
		Severe CS (SCS) & T3.5 & 48-63 kts  \\
		
		Very Severe CS (VSCS) & T4.0-T4.5 & 64-89 kts  \\
		
		Extremely Severe CS (ESCS) & T5.0-T6.0 & 90-119 kts  \\
		
		Super CS & T6.5 -T8.0 & $>$ 120 kts  \\
		\hline
	\end{tabular}
	\vspace{2mm}
	\caption{Different phases of CDs according to the IMD classification}
	\label{tab1}
	
\end{table}

CDs possess the potential to cause extensive damage due to their combination of severe winds, heavy rainfall, storm surges, and associated hazards. Let's discuss these factors separately:
\begin{enumerate}
	\item \textit{High-velocity winds:} CDs are characterized by extremely strong winds that can reach speeds well over 100 miles per hour (160 kilometers per hour). These powerful winds can uproot trees, damage buildings, and infrastructure, including roofs, windows, and walls. Flying debris propelled by the high winds can also pose a significant threat to both people and property.
	\item \textit{Heavy rainfall and flooding:} CDs are often accompanied by torrential rainfall. The intense precipitation can lead to flash floods and widespread flooding, overwhelming drainage systems and causing water levels to rise rapidly. This can result in extensive damage to homes, roads, bridges, and other infrastructure. Floodwaters can also contaminate water sources, leading to health risks and the destruction of crops and livestock.
	\item \textit{Storm surges:} One of the most devastating consequences of  CDs is the storm surge. A storm surge is a rapid rise in sea level caused by strong winds and low atmospheric pressure associated with the CD. As the CD approaches the coast, it can push a wall of water onto the land, causing coastal flooding and widespread destruction in low-lying areas. Storm surges can breach seawalls, erode beaches, and inundate coastal communities, leading to loss of life and significant property damage.
	\item  \textit{Tornadoes and water spouts:}  CDs often generate tornadoes and water spouts, which can cause localized but intense damage. These rotating columns of air can produce violent winds that can demolish structures, uproot trees, and create a path of destruction in their vicinity. Tornadoes associated with  CDs can form rapidly and pose an additional threat to areas already impacted by the CD's primary effects.
	\item \textit{Landslides and erosion:} The heavy rainfall associated with  CDs can saturate the soil, increasing the risk of landslides and soil erosion. Steep slopes and areas with weak soil conditions are particularly vulnerable due to heavy rainfall. Landslides can bury homes, roads, and infrastructure, causing casualties and hindering rescue and relief efforts. Erosion of coastal areas can lead to the loss of valuable land, including beaches and dunes that act as natural buffers against future storms.
\end{enumerate}

Numerous case studies have examined human and economic losses resulting from CDs in various regions, including China \cite{bib9}, Bangladesh \cite{bib11}, Odisha \cite{bib12,bib13}, and Andhra Pradesh \cite{bib10}. Our study focuses on the NIO basin, comprising the Bay of Bengal and the Arabian Sea. The Bay of Bengal is one of the six prominent regions worldwide prone to CDs, with the most devastating ones originating from this region. Historically, CDs from the Bay of Bengal occur predominantly between October and November. A concerning trend in the Bay of Bengal is the rising sea temperatures, contributing to the intensification of CDs near coastal areas. This warming trend creates a favorable environment for CD strengthening, amplifying risks and storm impacts. To address this trend and mitigate its consequences, continuous monitoring, early warning systems, and proactive measures are necessary to enhance disaster preparedness and response in vulnerable coastal regions. The most affected countries in the Bay of Bengal basin are India, Bangladesh, and Myanmar, facing significant impacts from CDs. Similarly, the Arabian Sea basin primarily affects western India and Pakistan, leading to socio-economic and environmental consequences in their coastal regions. 

In India, all thirteen coastal states and Union Territories experience the impact of CDs, with specific regions like Tamil Nadu, Andhra Pradesh, Odisha, West Bengal, and Gujarat being particularly vulnerable due to the higher frequency and intensity of these events. The Indian Meteorological Department (IMD) serves a crucial role in monitoring and forecasting cyclonic weather conditions in India, providing early warnings and assisting disaster management authorities.
 
Bangladesh heavily relies on its coastal areas along the Bay of Bengal, where the Sundarbans, a renowned mangrove forest and UNESCO World Heritage Site, supports diverse ecosystems and unique wildlife.

Myanmar's western coastline along the Bay of Bengal holds economic, environmental, and cultural significance, including sandy beaches, mangrove forests, estuaries, and deltaic regions. The Irrawaddy Delta, formed by the Irrawaddy River, is notable for its fertile agricultural lands and distinct ecosystem.

Pakistan's coastal line extends along the Arabian Sea, particularly in the provinces of Sindh and Balochistan, facilitating trade and commerce through major ports. Understanding the factors contributing to losses caused by CDs is crucial for prevention and mitigation efforts.

Inland flooding resulting from cyclonic rainfall \cite{bib3,bib4,bib5} and high wind speeds \cite{bib6,bib7} have been identified as significant contributors to losses in CD-affected areas. However, detailed studies on these losses are hindered by the lack of comprehensive data. This study aims to analyze recent statistical and meteorological data to assess the frequency of CDs in the NIO. By examining CD occurrences and resulting damage in the countries surrounding the NIO, the study seeks insights into the influence of these cyclonic events on the affected nations.

The structure of this paper includes Section \ref{sec2}, which describes the data and methods used, Section \ref{sec3}, which presents discussions on various aspects such as annual and seasonal variations on the generation of CDs, human losses, agricultural losses, and economic losses, and finally, Section \ref{sec4}, which concludes the paper.

\section{Data and Methods}\label{sec2}
\subsection{Data}
The IMD compiles an annual report \cite{bib8} through its Regional Specialized Meteorological Centre (RSMC) over the NIO region. This report contains crucial data regarding CD and serves as a valuable resource. It includes information such as the timing and location of CD generation and landfall, the tracks followed by the CD, the air pressure at the CD's center, the intensity and duration of the CD, landing sites, associated wind, rainfall intensities, Casualty, infrastructure losses, and agriculture losses. The IMD diligently collects comprehensive monitoring data on CDs in the NIO region every year. Furthermore, they meticulously record the extent of damage caused by these CDs, allowing for a comprehensive understanding of their impact on various aspects.

Since 1990, the IMD has consistently prepared and published an annual report through its RSMC over NIO. This report provides comprehensive documentation on various aspects, including the affected population, impacted areas and agricultural regions, casualties, damaged and destroyed houses, impaired infrastructure, direct economic losses, and agricultural economic losses resulting from meteorological disasters. These valuable datasets cover the period from 1990 to 2022, offering a wealth of information for researchers and policymakers who aim to analyze and mitigate the impact of cyclonic disturbances in the NIO region \cite{bib8}.

\subsection{Methods}
This study aims to conduct a thorough investigation to evaluate the societal impacts caused by natural disasters originating from NIOs. The primary objective is to assess the direct economic losses (which include the houses damage and agricultural losses) and human casualties resulting from these events. Utilizing statistical methods namely correlation, R-squared value, and linearly fitted equation, the study analyzes a dataset covering 32 years, allowing for a comprehensive approach that compares observations across different years. This approach leads to a deeper understanding of the patterns and trends concerning the social consequences of natural disasters over the studied time frame. In summary, through this in-depth investigation, the study aims to assess the societal impacts of CDs arising from NIOs and gain valuable insights into their long-term effects.

\section{ Discussion}\label{sec3}
\subsection{Trend in the generation of CDs over the NIO over a span of 32 years.}
The NIO basin, known for its dynamic nature among the seven global basins, has exhibited significant activity, as shown in Table \ref{tab2}. From 1990 to 2022, a total of 297 CDs were formed in this region, resulting in an average of 9 CDs per year. The year 2022 witnessed the highest number of CDs, with 15 occurrences, while 2012 had the lowest number, with only 4 instances. Figure 1 provides a visual representation of the variations in CD numbers in the NIO from 1990 to 2022.

Throughout the past few decades, the average annual CD count in the NIO was 7 in the 1990s, 8 in the 2000s and 2010s. However, in the 2020s, there was a slightly higher average of 11 CDs. This suggests an upward trend in the number of CDs generated in the NIO, emphasizing the notable changes observed in recent years. 

The numbers have shown fluctuations over time, and the current decade indicates an increase in CD occurrence compared to previous decades.

\begin{longtable}{|c|c|c|c|c|c|c|c|c|c|} 
	\captionsetup{justification=centering, skip=10pt} 
	\caption{From 1990 to 2022, a significant number of CDs were formed over the NIO. BB stands for Bay of Bengal, AS for Arabian Sea, LD for Land} \\
	\hline
	Year & Basin & D & DD & CS & SCS & VSCS & ESCS & SuCS & Total \\
	\hline
	&BB&6&2&0&2&0&0&0&10 \\
	1990&AS&0&0&0&0&0&0&0&0 \\
	&LD&0&0&0&0&0&0&0&0 \\
	\hline
	&Total&&&&&&&&10 \\
	\hline
	&BB&4&1&1&0&0&0&1&7 \\
	1991&AS&0&1&0&0&0&0&0&1 \\
	&LD&0&0&0&0&0&0&0&0 \\
	\hline
	&Total&&&&&&&&8 \\
	\hline
	&BB&0&3&3&1&0&0&0&7 \\
	1992&AS&0&1&3&0&0&0&0&4 \\
	&LD&0&0&0&0&0&0&0&0 \\
	\hline
	&Total&&&&&&&&11\\
	\hline
	&BB&0&2&1&0&0&0&0&3 \\
	1993&AS&1&0&0&1&0&0&0&2 \\
	&LD&0&0&0&0&0&0&0&0 \\
	\hline
	&Total&0&0&0&0&0&0&0&5 \\
	\hline
	&BB&1&2&0&1&0&0&0&4 \\
	1994&AS&0&0&0&1&0&0&0& 1 \\
	&LD&0&0&0&0&0&0&0&0 \\
	\hline
	&Total&&&&&&&&5 \\
	\hline
	&BB&1&4&0&2&0&0&0&7 \\
	1995&AS&0&0&1&0&0&0&0&1 \\
	&LD&0&0&0&0&0&0&0&0 \\
	\hline
	&Total&&&&&&&&8 \\
	\hline
	& BB&1&4&1&1&1&0&0&8 \\
	1997&AS&1&0&0&0&0&0&0&1 \\
	&LD&0&0&0&0&0&0&0&0 \\
	\hline
	& Total&&&&&&&&9 \\
	\hline
	&BB&0&3&0&1&2&0&0&6 \\
	1998&AS&0&1&1&1&1&0&0&4 \\
	&LD&1&0&0&0&0&0&0&1\\
	\hline
	&Total&&&&&&&&11\\
	\hline
	&BB&2&2&1&0&1&0&1&7 \\
	1999&AS&0&0&0&0&1&0&0&1 \\
	&LD&1&0&0&0&0&0&0&1 \\
	\hline
	&Total &&&&&&&&9 \\
	\hline
	&BB&1&1&2&0&2&0&0&6 \\
	2000&AS&0&0&0&0&0&0&0&0 \\
	&LD&1&0&0&0&0&0&0&1 \\
	\hline
	&Total&&&&&&&&7 \\
	\hline 
	&BB&2&0&1&0&0&0&0&3 \\
	2001&AS&0&0&2&0&1&0&0&3 \\
	&LD&0&0&0&0&0&0&0&0 \\
	\hline
	&Total&&&&&&&&6 \\
	\hline
	&BB&1&1&2&1&0&0&0&5 \\
	2002&AS&0&0&0&0&0&0&0&0 \\
	&LD&0&0&0&0&0&0&0&0 \\
	\hline
	&Total&&&&&&&&5 \\
	\hline
	&BB&2&2&0&1&1&0&0&6 \\
	2003&AS&0&0&0&1&0&0&0&1 \\
	&LD&0&0&0&0&0&0&0&0 \\
	\hline
	&Total&&&&&&&&7 \\
	\hline
	&BB&2&0&0&0&1&0&0&3 \\
	2004&AS&0&2&0&3&0&0&0&5 \\
	&LD&2&0&0&0&0&0&0&2 \\
	\hline
	&Total&&&&&&&&10 \\
	\hline
	&BB&2&3&4&0&0&0&0&9 \\
	2005&AS&2&0&0&0&0&0&0&2 \\
	&LD&1&0&0&0&0&0&0&1 \\
	\hline
	&Total&&&&&&&&12 \\
	\hline
	&BB&5&2&1&0&1&0&0&9 \\
	2006&AS&0&1&0&1&0&0&0&2 \\
	&LD&1&0&0&0&0&0&0&1 \\
	\hline
	&Total&&&&&&&&12 \\
	\hline
	&&BB&3&4&1&0&1&0&9 \\
	2007&AS&0&1&1&0&0&0&1&3 \\
	&LD&0&0&0&0&0&0&0&0 \\
	\hline
	&Total&&&&&&&&12 \\
	\hline
	&BB&1&2&3&0&1&0&0&7 \\
	2008&AS&1&1&0&0&0&0&0&2 \\
	&LD&1&0&0&0&0&0&0&1 \\
	\hline
	&Total&&&&&&&&10 \\
	\hline
	&BB&0&2&2&1&0&0&0& 5 \\
	2009&AS&2&0&1&0&0&0&0&3 \\
	&LD&0&0&0&0&0&0&0&0 \\
	\hline
	&Total&&&&&&&&8 \\
	\hline
	&BB&2&1&0&2&1&0&0&6 \\
	2010&AS&0&0&1&0&1&0&0&2 \\
	&LD&0&0&0&0&0&0&0&0 \\
	\hline
	&Total&&&&&&&&8 \\
	\hline
	&BB&2&2&0&0&1&0&0&5 \\
	2011&AS&1&2&1&0&0&0&0&4 \\
	&LD&1&0&0&0&0&0&0&1 \\
	\hline
	&Total&&&&&&&&10 \\
	\hline
	&BB&0&2&1&0&0&0&0&3 \\
	2012&AS&0&0&1&0&0&0&0&1 \\
	&LD&0&0&0&0&0&0&0&0 \\
	\hline
	&Total&&&&&&&&4 \\
	\hline
	&BB&3&0&1&1&3&0&0&8 \\
	2013&AS&0&1&0&0&0&0&0&1 \\
	&LD&1&0&0&0&0&0&0&1 \\
	\hline
	&Total&&&&&&&&10 \\
	\hline
	&BB&2&2&0&0&1&0&0&5 \\
	2014&AS&0&0&1&0&1&0&0&2 \\
	&LD1&0&0&0&0&0&0&0&1 \\
	\hline
	&Total&&&&&&&&8 \\
	\hline
	&BB&1&1&1&0&0&0&0&3 \\
	2015&AS&0&2&1&0&0&2&0&5 \\
	&LD&2&2&0&0&0&0&0&4 \\
	\hline
	&Total&&&&&&&&12 \\
	\hline
	&BB&1&2&3&0&1&0&0&7 \\
	2016&AS&2&0&0&0&0&0&0&2 \\
	&LD&1&0&0&0&0&0&0&0 \\
	\hline
	&Total&&&&&&&&10 \\
	\hline
	&BB&4&1&1&1&1&0&0&8 \\
	2017&AS&0&0&0&0&0&0&0&0 \\
	&LD&2&0&0&0&0&0&0&2 \\
	\hline
	&Total&&&&&&&&10 \\
	\hline
	&BB&3&2&1&2&1&0&0&9 \\
	2018&AS&1&0&0&0&1&2&0&4 \\
	&LD&1&0&0&0&0&0&0&1 \\
	\hline
	&Total&&&&&&&&14 \\
	\hline
	&BB&0&1&1&0&1&1&0&4\\
	2019&AS&2&1&1&0&2&1&1&8 \\
	&LD&0&0&0&0&0&0&0&0 \\
	\hline
	&Total&&&&&&&&12 \\
	\hline
	&BB&1&1&1&0&1&0&1&5 \\
	2020&AS&2&0&0&1&1&0&0&4 \\
	&LD&0&0&0&0&0&0&0&0 \\
	\hline
	&Total&&&&&&&&9 \\
	\hline
	&BB&3&1&2&0&1&0&0&7 \\
	2021&AS&1&0&0&1&0&1&0&3 \\
	&LD&0&0&0&0&0&0&0&0 \\
	\hline
	&Total&&&&&&&&10 \\
	\hline
	&BB&4&3&1&2&0&0&0&10 \\
	2022&AS&2&1&0&0&0&0&0&3 \\
	&LD&2&0&0&0&0&0&0&2 \\
	\hline
	&Total&&&&&&&&15

	\label{tab2}
\end{longtable}

\begin{figure}[h]
	\includegraphics[width=1\textwidth]{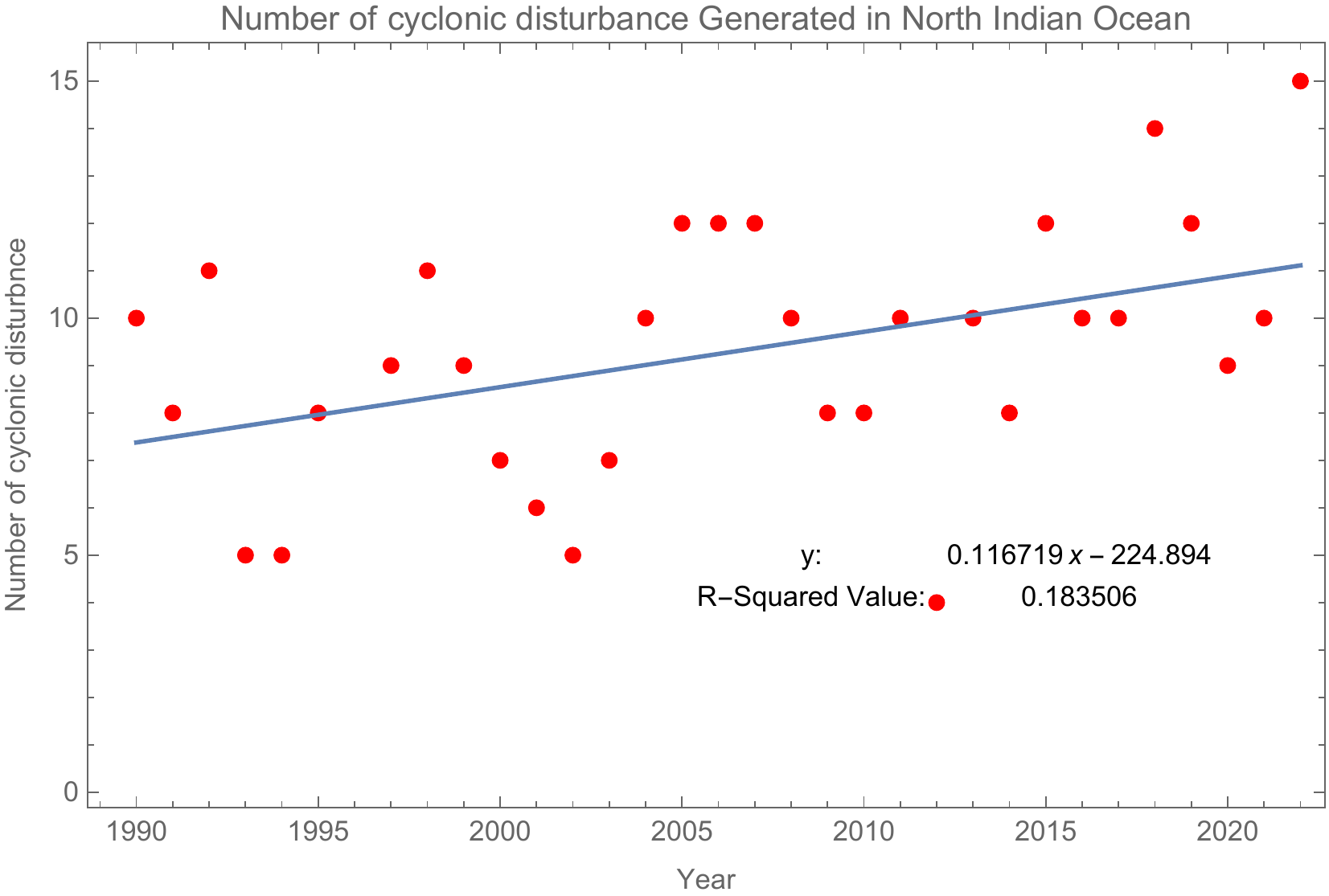}
	\label{fig1}
	\caption{The occurrence frequency of CDs generated in the NIO between 1990 and 2022 is depicted by the red dots, while the trend is represented by the solid blue line}
\end{figure}

\subsection{Seasonally trend in the generation of CD over the NIO}
Table \ref{tab3} presents insights on the average distribution of CDs generated during different months. The peak month for CD generation is November, with an average of 18.28 occurrences. Following closely behind is October, with an average of 16.79 CDs. May and June both exhibit an average of 11.56 CDs. On the other hand, February and March have the lowest average numbers of generated CDs, with a value of 0.7.

Of notable significance are October, November, and December, which collectively contribute to nearly half of all CDs formed in the NIO. Weather conditions in these months play a crucial role in CD generation, emphasizing the need for heightened vigilance and preparedness measures to mitigate potential impacts. In contrast, January, February, and March experience infrequent occurrences of CDs.
\begin{table}[h]
	\centering
	\caption{The average count of CDs generated in various months}
	\begin{tabular}{|c|c|}
		
		\hline
		Month& Generated CD ($\%$) \\
		\hline
		January & 1.11 \\
		Feburary & 0.7 \\
		March & 0.7 \\
		April & 1.49 \\
		May &11.56 \\
		June & 11.56 \\
		July & 7.46 \\
		August & 11.19 \\
		September & 7.83 \\
		October & 16.79 \\
		November & 18.28 \\
		December & 11.19 \\
		\hline
	\end{tabular}

\label{tab3}
\end{table}
\subsection{Casualty analysis}
\begin{figure}[h]
	\includegraphics[width=0.9\textwidth]{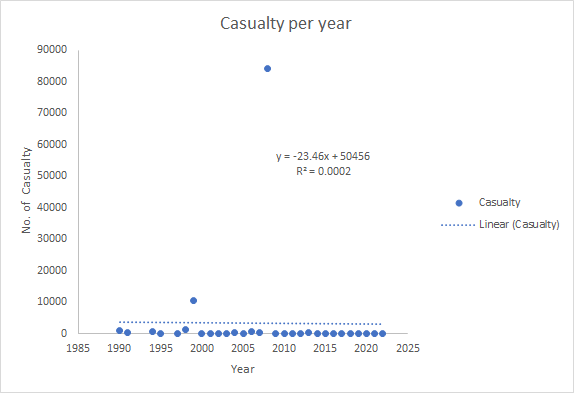}
	\caption{The casualties caused by CDs in India from 1990 to 2022 are represented by data points (dot), and the trend is depicted by a dashed line}
	\label{fig2}
\end{figure}

By utilizing historical data from 1990 to 2022, we can examine the cumulative number of casualties caused by CD disasters over a 32-year span, which totaled 101,021 individuals. On average, this equates to around 3,157 casualties per year attributed to these events. The data reveals a slight decrease in the number of casualties resulting from CDs since 2015, as shown in Figure \ref{fig2}. However, it is crucial to acknowledge the existence of data gaps between 1990 and 2022, which may affect the accuracy and comprehensive understanding of the situation.

There exists a negative correlation between the occurrence of CDs in the NIO and the resulting casualties. The decline in casualties can be attributed to the government's prioritization of people's well-being and their intensified efforts in disaster prevention and mitigation since the 2010s. These efforts aim to minimize the loss of human life during natural disasters, emphasizing the significance of reducing such casualties.

The year 2008 witnessed the highest number of casualties, with 84,113 fatalities, followed by 10,539 deaths in 1999 and 1,290 deaths in 1998. Particularly, the year 2008 stands out as the deadliest since 1990, with 10 CDs occurring in the NIO. Among these incidents, the Very Severe Cyclonic Storm "NARGIS" had the most devastating impact, claiming the lives of 84,000 individuals. NARGIS formed over the Bay of Bengal from April 27 to May 3, 2008, causing extensive destruction, including the sinking of vessels, destruction of houses, bridges, roads, and other infrastructure. Severe flooding, landslides, and mudslides were also triggered. Remarkably, NARGIS maintained its Very Severe Cyclonic Storm intensity for approximately 12 hours even after making landfall.

Table \ref{tab4} summarizes the top five deadliest CDs recorded between 1990 and 2022. These five events account for approximately $95\%$ of the total casualties associated with CDs during the specified period.

\begin{table}[h]
	\centering
	\caption{ Five most catastrophic CD between 1990 and 2022}
	\begin{tabular}{c|c|c|c|c}
		\hline
		Rank&Name&Year&Category&Casualty \\
		\hline
		1&Nargis&2008&VSCS&84,000 \\
		2&October &1999& SuCS&9887\\
		3&June 1998&1998&VSCS&1173\\
		4&May 1990&1990&SuCS&928\\
		5&May&1999&VSCS&454\\
		\hline
	\end{tabular}

\label{tab4}
\end{table}

\subsection{Analysis of Economic Loss}
\textbf{Direct Economic loss :}

\begin{itemize}
	\item \textit{Infrastructure Damage:} CDs can cause severe damage to critical infrastructure such as roads, bridges, power grids, and communication networks. The cost of repairing and rebuilding infrastructure can be substantial.
	\item \textit{Property Damage:} Cyclonic winds and storm surges can lead to the destruction or severe damage of residential, commercial, and industrial buildings. The financial burden of rebuilding or repairing properties contributes to economic losses.
	
	\item \textit{Human Casualties:} Cyclonic rainfall, cyclonic flood, and storm surges have the potential to cause human casualties, resulting in profound sorrow and suffering for affected communities while also triggering significant economic ramifications. The consequences of human lives lost extend beyond the immediate aftermath of a CD, impacting various economic aspects. This includes reduced investment, diminished tourism, and decreased business activities in the affected regions. As a result, the overall economic growth of the area may be hindered, necessitating substantial resources for long-term recovery and reconstruction efforts. The loss of human lives not only has a devastating emotional impact but also carries long-lasting economic implications for the affected region. 
	\item  \textit{Disrupted Economic Activities:} CDs can disrupt economic activities, particularly in sectors such as agriculture, fisheries, tourism, and transportation. Crop and livestock losses, disruption of supply chains, reduced tourist arrivals, and halted trade can all have significant economic consequences.
	
\end{itemize}

\textbf{Indirect Economic loss :}

\begin{itemize}
	\item \textit{Loss of Productivity:} Disrupted economic activities and damaged infrastructure can lead to a loss of productivity in various sectors. Businesses may struggle to operate, leading to reduced production, employment, and income generation.
	
	\item  \textit{Income Loss:} CD-related disruptions can result in job losses, reduced working hours, and income reduction for individuals and businesses. This loss of income can have long-term implications for individuals and the overall economy.
	
	\item  \textit{Increased Expenditure:} Governments often bear the financial burden of CD response and recovery efforts. Increased expenditure on emergency response, relief measures, and infrastructure rehabilitation can strain public finances and divert resources from other development priorities.
\end{itemize}

This study's primary objective is to analyze direct economic losses, specifically those incurred in agriculture, as well as the damage caused to infrastructure, such as houses. We solely focus on these aspects, investigating the extent of their economic impact. Our analysis aims to comprehensively understand the direct economic losses (argiculture and infrastructure damage), shedding light on the overall consequences.

\subsubsection{Demographical Analysis of Damaged House }
Over a span of 32 years, from 1990 to 2022, a thorough examination of historical data highlights a significant finding: the total count of houses damaged due to CD disasters has reached an astonishing figure of 5,098,257. On an annual average, approximately 159,320 houses experienced the consequences of these disasters. Analysis of the data reveals a gradual increase in the number of damaged houses, indicating the escalating impact of these disturbances (refer to figure 3). However, it is important to note that the presence of data gaps between 1990 and 2022 might influence the overall accuracy and our comprehensive understanding of the situation. Moreover, a positive correlation is observed between the occurrence of CD disasters over the NIO and the number of houses affected. Notably, the highest occurrence of damaged houses was recorded in 1999 (1,659,356), followed by 2008 (745,764) and 2019 (621,732).

\begin{figure}[h]
	\includegraphics[width=\textwidth]{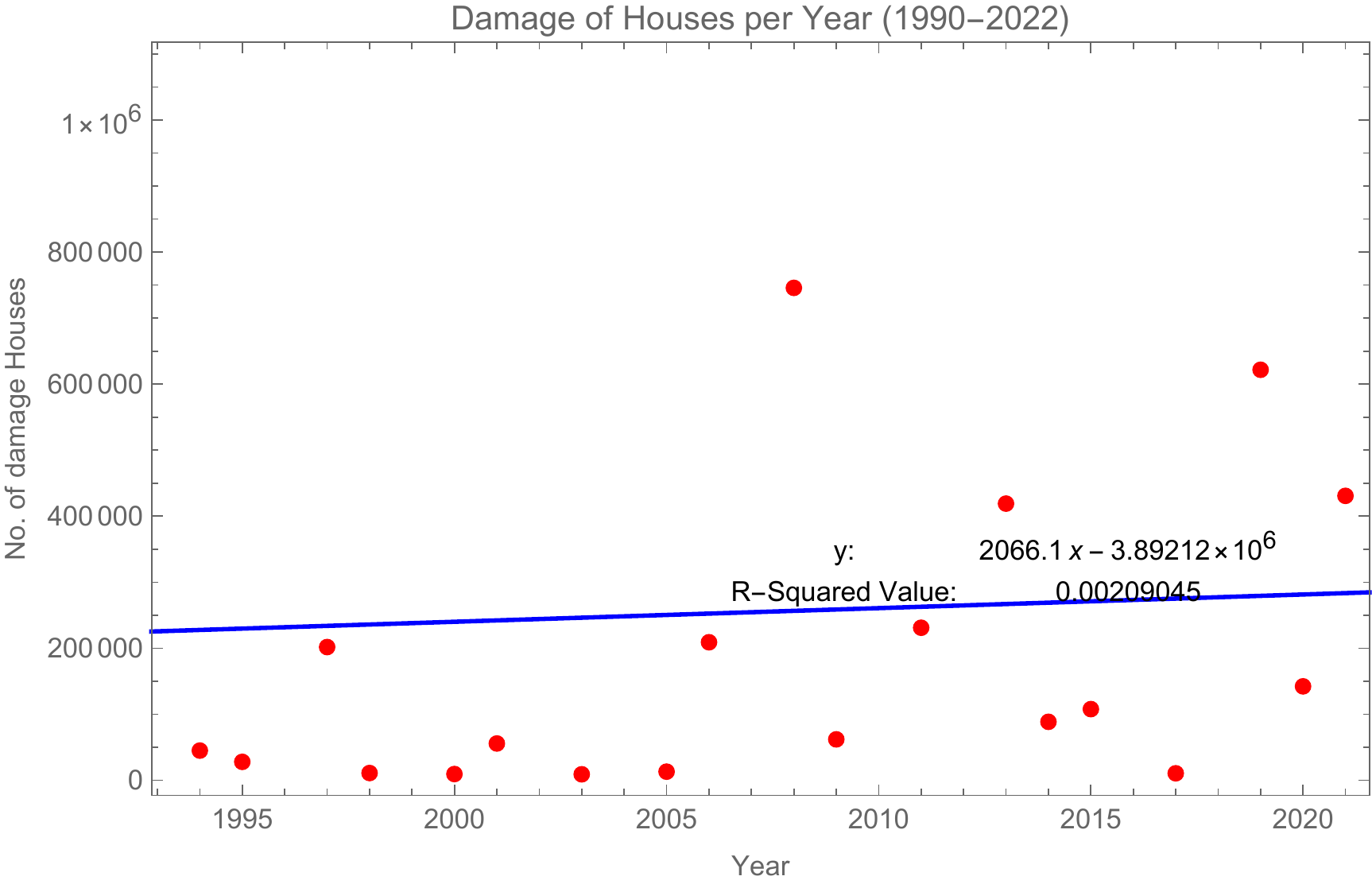}
	\caption{Number of damage houses due to CD between 1990 and 2022 is depicted by the red dots, while the trend is represented by the solid blue line}
	\label{fig3}
\end{figure}

\subsubsection{Demographical Analysis of Agriculture Loss }
\begin{figure}[h]
	\includegraphics[width=\textwidth]{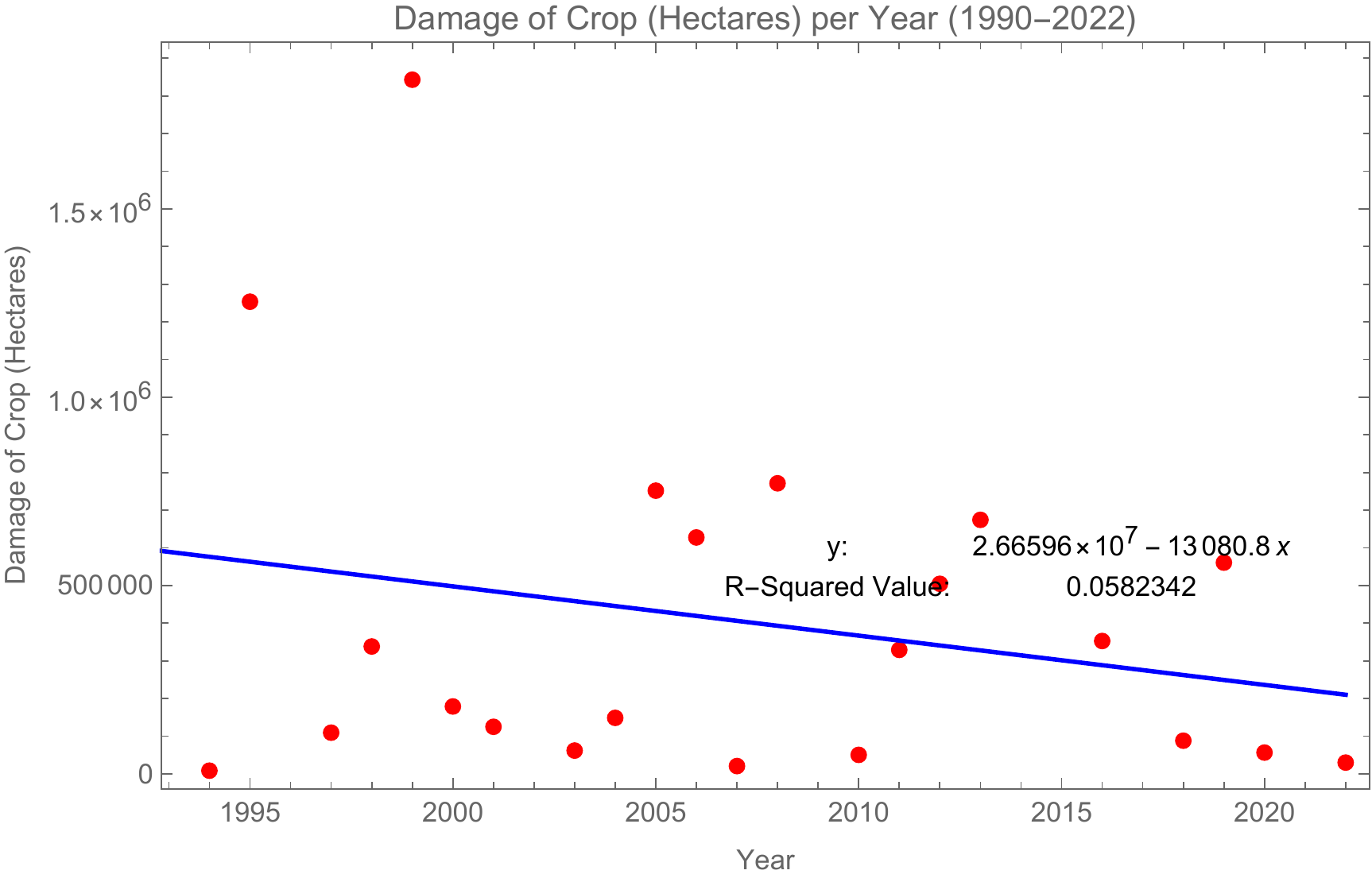}
	\caption{ Damage of Crop (Hectares) due to CD between 1990 and 2022 is depicted by the red dots, while the trend is represented by the solid blue line}
	\label{fig4}
\end{figure}

Based on a thorough analysis of historical data spanning from 1990 to 2022, it is evident that disasters caused by CDs have had a significant and detrimental effect on agriculture, resulting in substantial losses within the sector. The cumulative data over a span of 32 years reveals that a staggering 8,886,447 hectares of agricultural land have been affected by these disturbances, averaging approximately 277,701 hectares per year. A noteworthy observation is the downward trend in agricultural losses, indicating a progressive increase in the magnitude of these disruptions (see fig. \ref{fig4}).

However, it is important to acknowledge the presence of data gaps within the 1990-2022 timeframe, which may impact the overall accuracy and our comprehensive understanding of the situation. Despite this limitation, it is apparent that the incidence of CDs in the North Indian Ocean (NIO) region correlates negatively with agricultural loss. The highest agricultural losses were recorded in 1990, affecting 1,843,000 hectares, followed by 1995 with 1,253,653 hectares, and 2008 with 771,458 hectares.

As mentioned earlier, it is crucial to recognize that factors such as agriculture loss and infrastructure damage play a pivotal role in influencing the economic consequences resulting from CDs. Notably, the trends observed in agricultural and infrastructure losses suggest an overall upward economic trend from 1990 to 2022. Nevertheless, it is imperative to note the existence of data gaps within the available economic records during this period, which restricts our ability to obtain a complete and accurate understanding of the situation. These data gaps present a challenge to our comprehensive assessment of the overall economic impact caused by the CDs.
\section{Conclusion}\label{sec4}
The IMD has publish the annual report \cite{bib8} of the period from 1990 to 2022, providing valuable data on CDs in the NIO. This data encompasses important details such as the timing and location of CD generation and landfall, the tracks followed by the CD, the air pressure at the CD's center, the intensity and duration of the CD, landing sites, associated wind, rainfall intensities, casualty, infrastructure losses, and agriculture losses. It serves as a crucial foundation for analyzing various characteristics associated with CDs. Over the span of 32 years, a total of 297 CDs have been documented, averaging around 9 disturbances per year. Notably, there has been a significant increase in the number of CDs generated during this period.

However, it is important to acknowledge that there are data gaps within the timeframe of 1990 to 2022, which may impact the overall accuracy and comprehensive understanding of the situation. In order to assess the impacts of CDs, this study specifically focuses on the direct economic losses, including agriculture and infrastructure loss (specifically houses), as well as casualties recorded between 1990 and 2022. Historical data reveals that there were a total of 101,021 casualties recorded over the course of 32 years, averaging approximately 3,157 casualties per year from CDs making landfall. Interestingly, there appears to be a slight decreasing trend in the number of casualties since 2015.

Among the CDs analyzed, the five most devastating ones accounted for approximately $95\%$ of the total casualties during the studied period. Furthermore, the analysis of the data shows a slight upward trend in economic losses during the study period. These findings can be attributed to the proactive measures taken by the government of the countries surrounding the NIO in recent years, as they have focused on enhancing natural disaster prevention and mitigation efforts.

\section*{Acknowledgment}
I would like to express my sincere gratitude to the IMD for their invaluable contribution in publishing the comprehensive report on CDs. Their dedication and commitment to providing accurate and insightful data have greatly contributed to our understanding of the social impacts of these events.
\section*{Declaration}
\subsection*{Conflict of Interest}
Conflict of interest is not declared by any of the authors.
\subsection*{Author Statement}
Research and manuscript preparation were done equally by each author.
\subsection*{Data Availability}
Readily accessible on the IMD's official website
\subsection*{Funding}
This work was supported by Delhi Technological University under roll number 2k21/PHDAM/04. First Author ``Monu yadav" has received research support from Delhi Technological Univeristy.
\bibliography{sn-bibliography}


\end{document}